\documentstyle[aps,pre,manuscript,epsf,epsfig]{revtex}

\begin{document}

\draft

\title{Class of self-limiting growth models in the presence of nonlinear
diffusion}

\author{Sandip~Kar, Suman~Kumar~Banik and 
Deb~Shankar~Ray{\footnote{Electronic-mail: pcdsr@mahendra.iacs.res.in}}
}

\address{Indian Association for the Cultivation of Science, Jadavpur,
Calcutta 700 032, India}

\date{\today}

\maketitle

\begin{abstract}
The source term in a reaction-diffusion system, in general, does not involve
explicit time dependence. A class of self-limiting growth models dealing
with animal and tumor growth and bacterial population in a culture, on the
other hand are described by kinetics with explicit functions of time. We
analyze a reaction-diffusion system to study the propagation of spatial front
for these models.
\end{abstract}



\section{Introduction}

Reaction-diffusion systems are ubiquitous in almost all branches of 
physics \cite{ld}, chemistry \cite{ire} and biology \cite{nfb,jdm,mkot}
dealing with population growth, fluid dynamics, pulse
propagation in nerves, chemical reactions, optical and other processes. The
basic equation describes the dynamics of a field variable $n(x,t)$, a
function of space and time in terms of a source term (also known as reaction
term) and a diffusion term. An important early endeavor in this direction is
the study of self-limiting growth models of which the most well-known is the
Fisher equation \cite{raf,kpp} which takes into account of a linear growth 
and a nonlinear
decay. The model and many of its variants have found wide applications both
from theoretical and experimental point of view \cite{jdm}. A notable feature 
of these
models is that the source or the reaction terms do not involve any explicit 
time dependence. On the other hand there are situations \cite{ref1,ref2,ref3}
where the source 
terms contain explicit functions of time which put a constraint on the growth
process in the long time limit. For example, the Gompertz growth 
\cite{ref1,ref2} is a model
used for study of growth of animals and tumors, where the growth rate is 
proportional to the current value, but the proportionality factor decreases
exponentially in time so that

\begin{mathletters}

\begin{equation}
\label{eq1a}
\frac{dn}{dt} = r n \exp (-\alpha t) \; \; ,
\end{equation}

\noindent
where $r$ and $\alpha$ are positive experimentally determined constants. 
Similarly another type of        
model proposed to analyze the growth of bacterial population in culture
\cite{ref3} is described by

\begin{equation}
\label{eq1b}
\frac{dn}{dt} = k n t \exp (-\beta t^2) \; \; .
\end{equation}

\end{mathletters}

\noindent
Again
$k$ and $\beta$ are positive constants required to fit the experimental data.
An important feature of these models is that unlike the logistic growth
process the asymptotic value of the density function $n$ depends on its
initial population.

Keeping in view of these experimental observations it is therefore worthwhile
to generalize the specific cases in terms of an explicit function of time
$\phi(t)$ such that we write

\begin{equation}
\label{eq2}
\frac{dn}{dt} = r n \phi(t)
\end{equation}

\noindent
where $r$ is a constant for the growth process and $\phi(t)$ may of the type
(i) $\phi(t)$ = 1 for  exponential growth (ii) $\phi(t) = \exp (-\alpha t)$ for
Gompertz growth (iii) $\phi(t) = t \exp (-\beta t^2)$ for bacterial growth, etc.

The object of the present paper is to study a reaction-diffusion system with  
a reaction term describing a class of self-limiting growth processes 
(\ref{eq2}). Since in many living organisms concentration dependent 
diffusivity \cite{jdm,mkot,ns,ao,mrr,hm,mmkk} has been found to be
essential to the modeling of reaction-diffusion systems we investigate the
interplay of this nonlinear diffusion and self-limiting growth process in the
dynamics. We show that the model and its variant with a finite memory
transport \cite{cc,kph2,th,vm1,wh2,mhk,ga,sf,jms,rdb}
admit of exact solutions. The dependence of the rate of spread of
the wave front on various parameters is explored.

\section{The reaction-diffusion system}

We consider a reaction-diffusion system with a source term describing
self-limiting growth and with a nonlinear diffusion term in the following 
form:

\begin{equation}
\label{eq3}
\frac{\partial n(x,t)}{\partial t} = r n \phi(t)
+\frac{\partial}{\partial x} D n \frac{\partial n}{\partial x}
\end{equation}

\noindent
where D is the diffusion coefficient. Our primary aim in this section is to
provide an exact solution of Eq.(\ref{eq3}). To this end we first make use of the
following transformation

\begin{equation}
\label{eq4}
n(x,t) = \tilde{u}(x,t) \exp \left ( r \int_0^t \phi(t') d t' \right )
\end{equation}

\noindent
in Eq.(\ref{eq3}) to obtain

\begin{equation}
\label{eq5}
\frac{\partial \tilde{u}(x,t)}{\partial t}
= D \exp \left ( r \int_0^t \phi(t') d t' \right )
\frac{\partial}{\partial x}
\left \{ \tilde{u}
\frac{\partial \tilde{u} }{\partial x} \right \} \; \; .
\end{equation}

\noindent
We now introduce the scaled time variable $\tau$ as

\begin{mathletters}

\begin{equation}
\label{eq6a}
\tau = D \int_0^t f(t') d t' \equiv G(t) \; \; \; (say)
\end{equation}

\noindent
where

\begin{equation}
\label{eq6b}
f(t) = \exp [ r \int_0^t \phi(t') d t' ] \; \; .
\end{equation}

\end{mathletters}

This reduces Eq.(\ref{eq5}) to the following form

\begin{equation}
\label{eq7}
\frac{\partial u(x,\tau)}{\partial \tau}
= \frac{\partial}{\partial x}
\left \{ u(x,\tau) \frac{\partial u(x,\tau)}{\partial x} \right \}
\end{equation}

\noindent
with $\tilde{u}(x,t) \equiv \tilde{u}(x,G^{-1}(\tau)) = u(x,\tau)$
where time $t$ has been expressed as an inverse function $G^{-1}(\tau)$
according to Eq.(\ref{eq6a}-\ref{eq6b}).

Eq.(\ref{eq7}) is the well-known Boltzmann nonlinear diffusion equation
\cite{ld,jc}.
Now subject to the initial condition of a unit point source at the origin,

\begin{equation}
\label{eq8}
n(x,0) = \delta (x) = \tilde{u}(x,0) = u(x,0)
\end{equation}

\noindent
we solve Eq.(\ref{eq7}) under the following boundary conditions
 
\begin{equation}
\label{eq9}
\lim_{x \rightarrow \pm \infty} u(x,\tau) = 0 \; \; \; \; \forall \tau > 0
\end{equation}
and

\begin{equation}
\label{eq10}
\int_{-\infty}^{+\infty} u(x,\tau) d x = 1 \; \; \; \; \forall \tau > 0
\; \; .
\end{equation}

Next we seek the similarity solution of the nonlinear diffusion equation (\ref{eq7}).
We make use of the well-known similarity transformation 
\cite{ld,mkot,jc,gb}

\begin{equation}
\label{eq11}
u = \tau^{-1/3} v(z) \; \; {\rm and} \; \;  z = x \tau^{-1/3}
\end{equation}

\noindent
in Eq.(\ref{eq7}) to obtain

\begin{equation}
\label{eq12}
3 \frac{d}{dz} \left ( v \frac{dv}{dz} \right )
 + v + z \frac{dv}{dz} = 0
\end{equation}

\noindent
On integration Eq.(\ref{eq12}) yields

\begin{equation}
\label{eq13}
3 \left ( v \frac{dv}{dz} \right ) + z v = 0
\end{equation}

\noindent
Since we are interested in the symmetric solutions with $v'(0) = 0$ we have
put the integration constant zero in going from Eq.(\ref{eq12}) to 
(\ref{eq13}). On further integration Eq.(\ref{eq13}) results in the solution

\begin{mathletters}

\begin{equation}
\label{eq14a}
\begin{array}{lll}
v(z) & = & (A^2-z^2)/6 \; \; \; \; |z| < A \\
& = & 0 \; \; \; \; |z| > A
\end{array}
\end{equation}
    
\noindent
where $A$ is a constant which can be determined from the condition 
(\ref{eq10}) to obtain

\begin{equation}
\label{eq14b}
A = \left (9 / 2 \right )^{1/3}
\end{equation}

\end{mathletters}
Therefore the solution of Eq.(\ref{eq7}) in $x$ and $\tau$ is given by

\begin{eqnarray}
u(x,\tau) & = & \frac{1}{6 \tau} \left [ A^2 {\tau}^{2/3} - x^2 \right ]
\; \; |x| < A {\tau}^{1/3} \nonumber\\
& = & 0 \; \; \; \; \; \; \; \; \; \;
|x| > A {\tau}^{1/3}
\label{eq15}
\end{eqnarray}

It is interesting to note that by virtue of the relations 
(\ref{eq6a}-\ref{eq6b}) $\tau$ is
dependent on $r$ and $\phi(t)$ which control the growth and self-limiting
factors, respectively of the source term. This implies that the shock-wave
like behaviour with propagating wave-front at $x = x_f = A {\tau}^{1/3}$
as evident from the similarity solutions (\ref{eq15}) 
critically depends on the reaction terms. Specifically, 
the wave front propagates in the medium with speed

\begin{equation}
\label{eq16}
\frac{dx_f}{dt} = \frac{1}{3} {\left ( \frac{9 D}{2} \right )}^{1/3}
f(t) {\left [ \int_0^t f(t') d t' \right ]}^{-2/3}
\end{equation}

\noindent
where $f(t)$ is given by (\ref{eq6a}) and in turn depends on the functional form
of $\phi(t)$.

We now consider two specific cases to illustrate the spatial propagation
patterns.

(i) $\phi(t)$ = 1

For a constant value of $\phi$ the model suggests
an exponential growth. The relation (\ref{eq6a}) 
in this case can then be utilized to
obtain $f(t) = \exp (r t)$ so that $\tau = (D/r) [\exp (r t) - 1]$.
Putting this expression for $\tau$ in the solution (\ref{eq15}) 
we have after using Eq.(\ref{eq4})

\begin{equation}
\label{eq17}
n (x,t) = \frac{
\left [ A^2 \left \{ (D/r) \left ( \exp (rt) - 1 \right ) \right \}^{2/3}
\right ] - x^2
}{
(6D/r) [ \exp (rt) - 1] \exp (-rt)
}
\end{equation}

\noindent
This solution clearly has a sharp wave-front at $x_f = A {\tau}^{1/3}$ which
propagates at a speed

\begin{equation}
\label{eq18}
\frac{dx_f}{dt} = \frac{1}{3} A \left (D r^2 \right )^{1/3} \exp (r t)
\left (\exp (r t) - 1 \right )^{-2/3}
\end{equation}

To illustrate the spatial propagation of the population $n(x,t)$ in time we
plot in Fig-1 the spatial shock-wave like patterns for $r = 1.0$ and 
$D = 1.0$. 
It is apparent that the sharply peaked distribution at $t = 0$ starts
spreading relatively slowly with peak at $x = 0$ diminishing with time upto a
period $t = 0.1$ . Beyond this time the spatial growth of population becomes
comparatively large and it diverges due to the combined effect of exponential
growth and nonlinear diffusion. For a much lower growth rate $(r = 0.001)$,
however, the population spreads monotonically due to the nonlinear diffusion
which overwhelms the effect of growth process. This is evident in Fig-1(b).

(ii) $\phi(t) = t \exp (-\beta t^2)$

With the above expression for $\phi (t)$ for bacterial self-limiting growth
we obtain from (\ref{eq6a}-\ref{eq6b})

\begin{equation}
\label{eq19}
f(t) = \exp [ (-r/2 \beta) (\exp(-\beta t) - 1)]
\end{equation}

\noindent
and

\begin{equation}
\label{eq20}
\tau = D \exp(r /2 \beta) \int_0^t \exp [ (-r/2 \beta) \exp (-\beta t' )] d t'
\end{equation}

\noindent
By defining $z = (r/2 \beta) \exp (-\beta t)$ the above expression can be
reduced to the following form

\begin{equation}
\label{eq21}
\tau = -D \frac{\exp(r / 2 \beta)}{ \beta}
\int_{(r/2 \beta)}^{(r/2 \beta) \exp (-\beta t)} \frac{\exp(-z)}{z} d z
\end{equation}

\noindent
The integral in (\ref{eq21}) can be put into a standard form with the help of
$Ei$-function \cite{gr} so that $\tau$ can be expressed as

\begin{equation}
\label{eq22}
\tau = D \frac{\exp (r/2 \beta)}{\beta}
[Ei \left (-r/2 \beta \right )
- Ei \left ((-r/2 \beta) \exp (-\beta t)\right )]
\end{equation}

\noindent
The corresponding density $n(x,t)$ and the speed of the wave front
$dx_f/dt$ at $x_f$ are given by


\begin{equation}
\label{eq23}
n(x,t) = \frac{A^2 \left (
 D \exp (r/2 \beta)/\beta \right )^{2/3}
\left [
Ei \left ( - r/2 \beta \right ) - 
Ei \left ( - (r/2 \beta) \exp (-\beta t) \right ) \right ]^{2/3} - x^2 
}{
(6 D/\beta)
\left [Ei (-r/2 \beta) - Ei ((-r/2 \beta) \exp (-\beta t))\right ]
\exp [(r/2 \beta) \exp (-\beta t)]}
\end{equation}

\noindent
and

\begin{equation}
\label{eq24}
\frac{dx_f}{dt} = A \left ( D \frac{\exp (r/2 \beta)}{\beta}\right )^{1/3}
\frac{d}{dt} \left [Ei (-r/2 \beta) - Ei ((-r/2 \beta) \exp (-\beta t))
\right]^{1/3}
\end{equation}


\noindent
respectively.

In Fig-2(a,b) we show the shock-wave like spread of population by plotting
$n(x,t)$ vs $x$ for several values of time for $D = 1$ and $r = 1$ . Since
$\beta$ puts a limit to the growth at large time the peak of $n(x,t)$ at
$x = 0$ as shown in Fig-2(a) ($\beta =0.1$) does not increase too much as
compared to the earlier case considered in Fig-1(a). It has been observed 
that for a unique value of $\beta \geq 1.0$ there is a monotonic decrease
in the peak population $n (x,t)$ at $x=0$. For smaller values of $\beta$ 
(Fig-2(b))
the spread is similar to that in Fig-1(a). In Fig-2(c) we exhibit the spatial 
front propagation for several values of growth rate $r$ at a time $t = 1.0$
keeping $D = 1$ and $\beta = 0.01$. It is apparent that with increase of $r$
the reaction dominates over diffusion so that the peak
population at $x =0$ increases compared to spreading.

\section{Effect of finite memory transport}

We now generalize the proposed reaction-diffusion model to include the effect
of finite memory transport. It has been observed that an animal's movement
at a particular instant of time often depends on its motion in the immediate
past. This results in a delay in population flux, or a memory in the diffusion
coefficient. A number of attempts have been made in the recent literature 
\cite{cc,kph2,th,vm1,wh2,mhk,ga,sf,jms,rdb} to 
analyze the delayed population growth in several models and related context
in heat conduction and transport processes. To consider a finite memory in the 
present model we modify the nonlinear diffusion term in Eq.(\ref{eq3}) to the following
form:

\begin{eqnarray}
\label{eq25}
& & \frac{\partial n(x,t)}{\partial t} =  r n(x,t) \phi(t) +
\nonumber \\
& &  \frac{\partial}{\partial x}
\left [D \gamma \int_0^t \exp[-\gamma (t - \tau)] n(x,\tau)
\frac{\partial n(x,\tau)}{\partial x} d \tau \right ]
\end{eqnarray}

\noindent
where $\gamma$ refers to the inverse of relaxation time. The population flux takes
into account of the relaxation effect due to the delay of the particles in
adopting a definite direction of propagation. Differentiating both sides of the 
above equation with respect to $t$ and using it again we obtain

\begin{equation}
\label{eq26}
\frac{{\partial}^2 n}{\partial t^2} = (r \phi - \gamma)
\frac{\partial n}{\partial t} + (r \dot{\phi} + r \phi \gamma) n
+ \frac{\partial}{\partial x}
\left [ D \gamma n \frac{\partial n}{\partial x} \right ]
\end{equation}

In the limit of vanishing relaxation time i.e, $1/ \gamma 
\rightarrow$0 Eq.(\ref{eq26}) reduces to Eq.(\ref{eq3}). 
When memory effects are taken into account, the dispersal of the organisms
are not mutually independent. Hence the correlation between the
successive movement of the diffusing particles results in a delay
in the transport. Thus Eq.(\ref{eq26}) is a typical form of a delayed
transport equation.

We now consider a specific case $\phi(t) = 1$. Substitution of the traveling
wave form $N(z)$ $(= n(x,t))$ with $z = x + c t$ satisfies

\begin{equation}
\label{eq27}
c^2 \frac{{\partial}^2 N}{\partial z^2} = c (r - \gamma)
\frac{\partial N}{\partial z} + r \gamma N + D \gamma
\frac{\partial}{\partial z} \left ( N \frac{\partial N}{\partial z}\right )
\end{equation}

\noindent
where $c$ is the speed of the traveling wave to be determined.

We now consider the trial solution of Eq.(\ref{eq27}) of the form $N(z) =
N_0 \exp (s z^b)$ subject to the initial condition that at $z = 0$,
$N = N_0$, where $s$ and $b$ are positive constants to be determined.
Substitution of this solution in Eq.(\ref{eq27}) yields the following relation

\begin{eqnarray}
& & [ c^2 s^2 b^2 z^{2(b-1)} + c^2 s b (b-1) z^{(b-2)} 
- c s b (r-\gamma) z^{(b-1)} \nonumber \\
& & - r \gamma ] \exp (s z^b) 
- D \gamma N_0 s b [ 2 s b z^{2(b-1)} + (b-1) z^{(b-2)}]
\nonumber \\
& & \times \exp (2 s z^b) \equiv L(z) = 0
\label{eq28}
\end{eqnarray}

\noindent
For $L(z) =0$, for all z, the coefficients of $\exp (s z^b)$ and
$\exp (2 s z^b)$ within the square brackets must vanish identically.
For this the only acceptable solution for $b$ is $b = 1$. We obtain

\begin{mathletters}

\begin{equation}
\label{eq29a}
2 s^2 D \gamma N_0 = 0
\end{equation}

\noindent
and

\begin{equation}
\label{eq29b}
c^2 s^2 - c s (r - \gamma) - r \gamma = 0
\end{equation}

\end{mathletters}

\noindent
From the above two equations the solution for $s$ is given by


\begin{equation}
\label{eq30}
s = \frac{c [(1/ \gamma) - (1/r)] +
[ c^2 \left ( (1/ \gamma) - (1/r) \right )^2 + 4/r 
\left ( (c^2/ \gamma) + 2 D N_0 \right )]^{1/2}}
{2 \left [ \left (c^2 / \gamma r \right ) + 
\left ( 2 D N_0 / r \right ) \right ]}
\end{equation}


\noindent
In the limit of instantaneous relaxation, i.e, $1/ \gamma \rightarrow$0 
Eq.(\ref{eq30}) yields 

\begin{equation}
\label{eq31}
s = \frac{c \left [-1 + \left (1 + (D N_0 r/ c^2) \right )^{1/2} \right ]}
{ 4 D N_0}
\end{equation}

\noindent
Furthermore the above expression in the limit of weak diffusion 
$D \rightarrow$0 we obtain from Eq.(\ref{eq31}) after a Taylor expansion

\begin{equation}
\label{eq32}
s = \frac{r}{c}
\end{equation}

To determine the speed of the propagation of the wave front we now rearrange
the solution for $s$ in (\ref{eq30}) to obtain 

\begin{equation}
\label{eq33}
c = \frac{ (r - \gamma) + \left [ (r - \gamma)^2 +
4 (r \gamma - 2 s^2 D \gamma N_0) \right ]^{1/2}}
{2 s}
\end{equation}

\noindent
For real values of $c$, the quantity inside the square root must be positive,
which determines the minimum value of $c$ for $s = r/c$ [ Eq.(\ref{eq32}) ] as

\begin{equation}
\label{eq34}
c_{min} = \frac{2 r^2 D \gamma N_0}
{(r + \gamma)^2}
\end{equation}

Eq.(\ref{eq27}) therefore admits of an exact traveling wave like solution


\begin{equation}
N(z) = N_0 
\exp \left [ \frac{c (r - \gamma) + 
\left ( c^2 (r - \gamma)^2 + 4 r \gamma (c^2 + 2 D \gamma N_0) \right )^{1/2}}
{2 (c^2 + 2 D \gamma N_0)} \right ] z
\label{eq35}
\end{equation}


It is interesting to observe that the speed of the traveling wave front
not only depends on nonlinear diffusion and growth rate but also on the 
initial concentration and memory. 
A comparison of the solutions in this section and in the previous one
shows that (\ref{eq35}) does not reduces to Eq.(\ref{eq17}) in the
limit of vanishing relaxation time ($1/\gamma \rightarrow 0$) although
Eq.(\ref{eq26}) goes over to Eq.(\ref{eq3}) under this condition. This
is because of the fact that the nature of the partial differential equation
changes due to the inclusion of relaxation terms and also the boundary 
conditions for the shock wave like `diffusing solutions' (\ref{eq17})
are different for the travelling wave front solution (\ref{eq35}). The
nature of the two solutions are thus generically different.
We point out in passing that the dependence
on initial concentration on speed as shown in (\ref{eq34}) is rather an 
unusual feature in reaction-diffusion system.

\section{Conclusions}

In this paper we have analyzed a class of reaction-diffusion systems in
which the kinetic term describes the self-limiting growth processes of
the Gompertz type and is an explicit function of time. We have shown that the
model can be solved exactly to analyze the spatial front propagation problem.
To make the model more realistic we have included the effect of finite
relaxation to concentration-dependent diffusive processes. In view of the fact
that the source terms have their direct relevance on experimental measurement
on animal and tumor growth or bacterial culture we think that the solutions
discussed in this paper will be pertinent in the context of reaction-diffusion
systems, in general.

\acknowledgements
The authors are indebted to C.S.I.R. (Council of Scientific and Industrial
Research), Government of India, for financial support.


\begin{figure}
\caption{Evolution of spatial front in time for the model with
$\phi (t)$ = 1. 
(a) The population $n(x,t)$ is plotted against $x$
for different times using $r$ = 1.0 and $D$ = 1.0. 
(b) Same as in Fig.(1a) but for $r$ = 0.001 (units arbitrary).
}
\label{fig1}
\end{figure}

\begin{figure}
\caption{Evolution of spatial front in time for the model with
$\phi (t)$ = $t \exp(-\beta t^2)$. 
(a) The population $n(x,t)$ is plotted against $x$
for different times using $r$ = 1.0, $D$ = 1.0 and $\beta$ = 0.1. 
(b) Same as in Fig.(2a) but for $\beta$ = 0.01.
(c) The population $n(x,t)$ is plotted against $x$ at $t$ = 1.0 for
different $r$ using $D$ = 1.0 and $\beta$ = 0.01 (units arbitrary).
}
\label{fig2}
\end{figure}

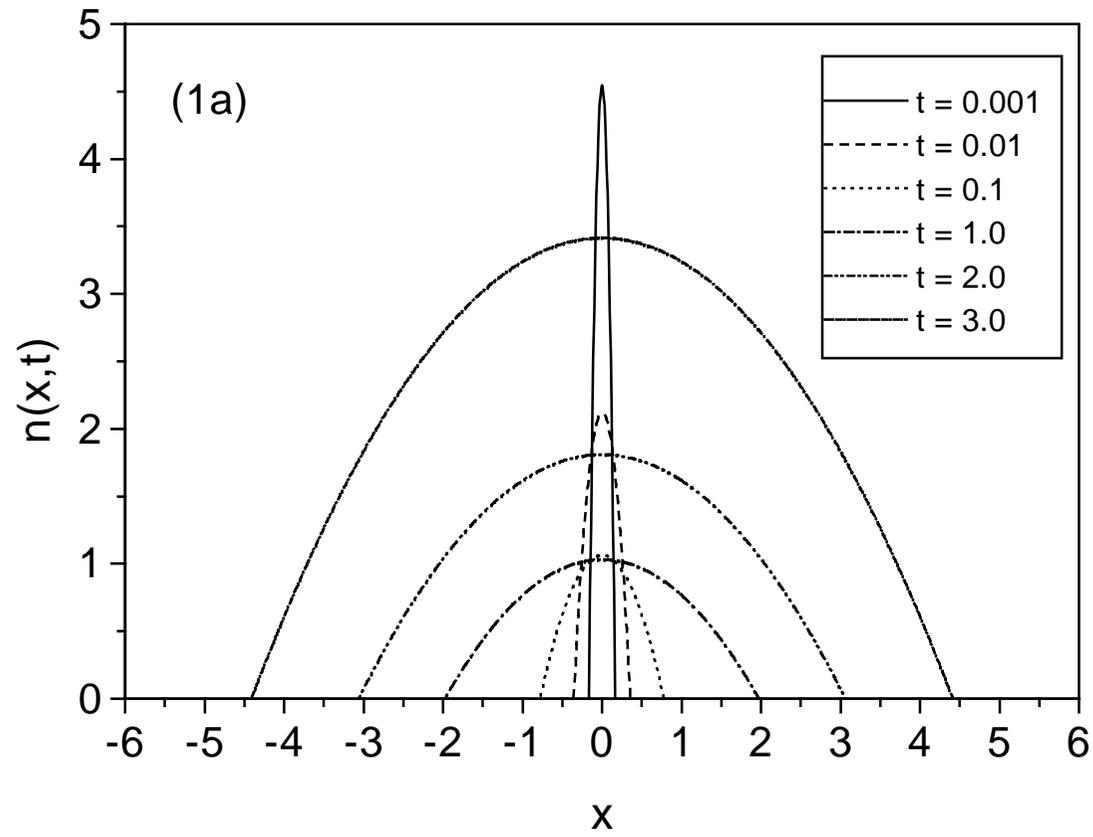
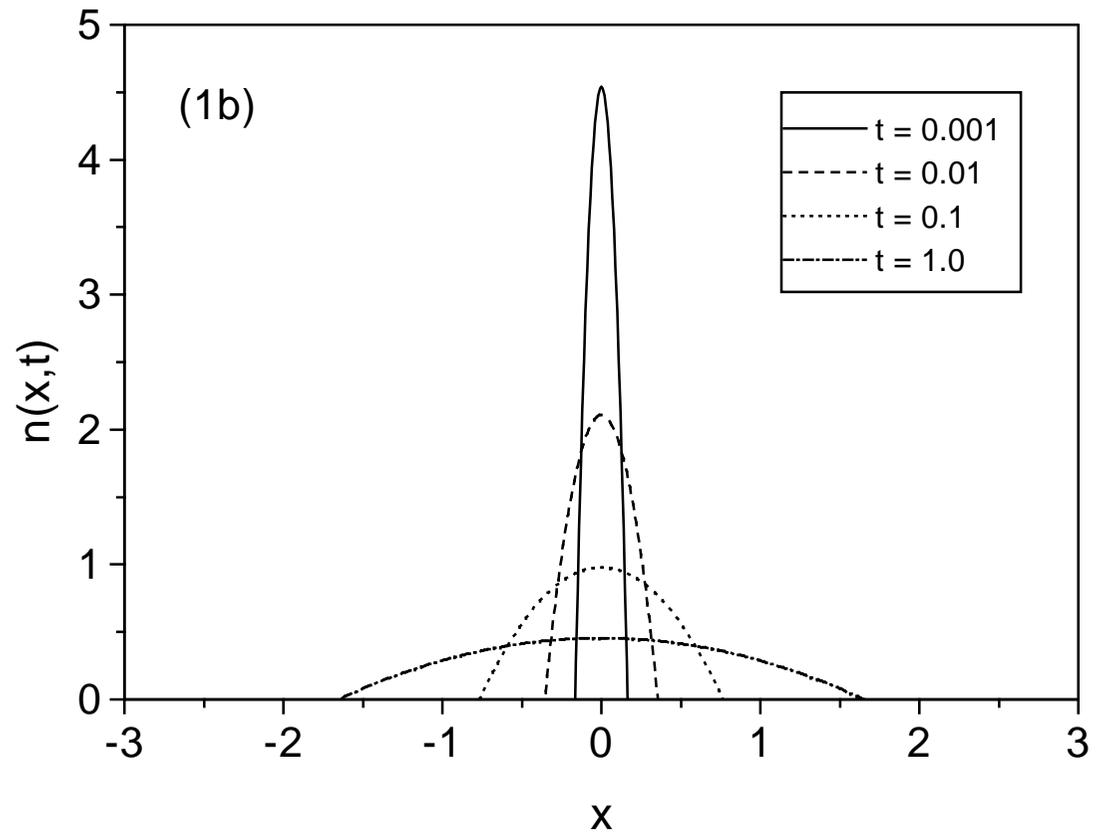
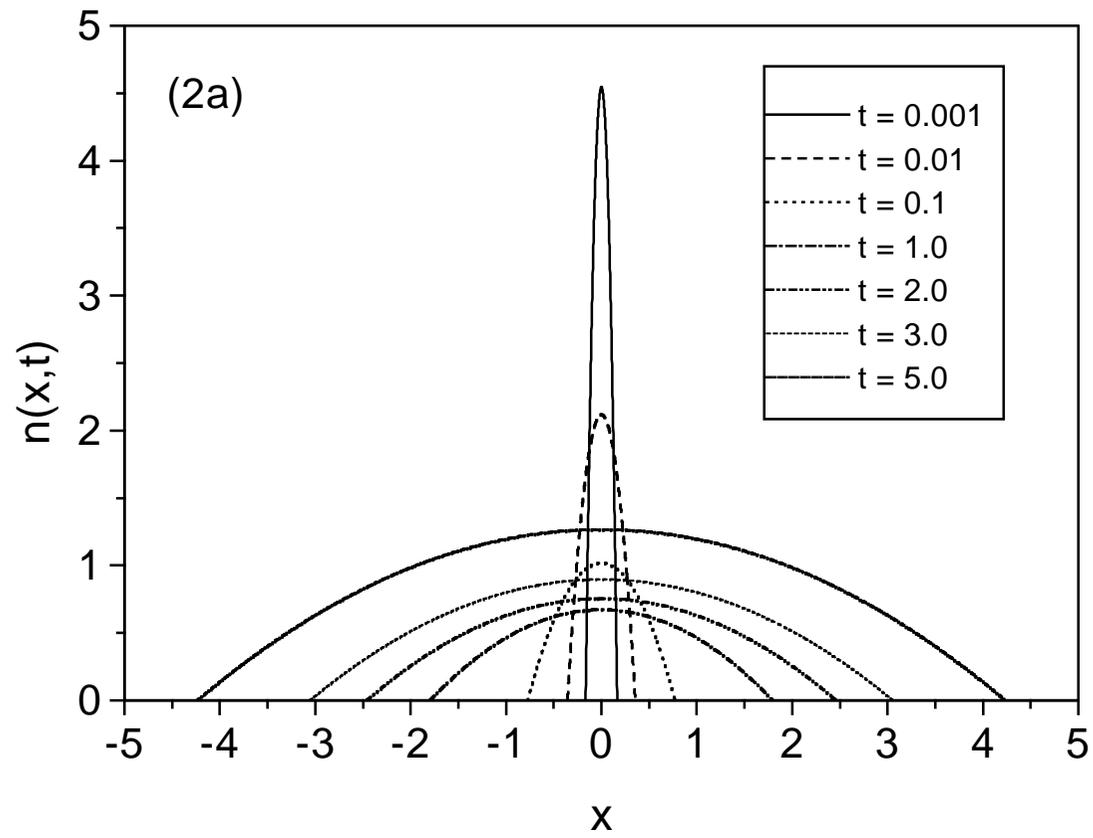
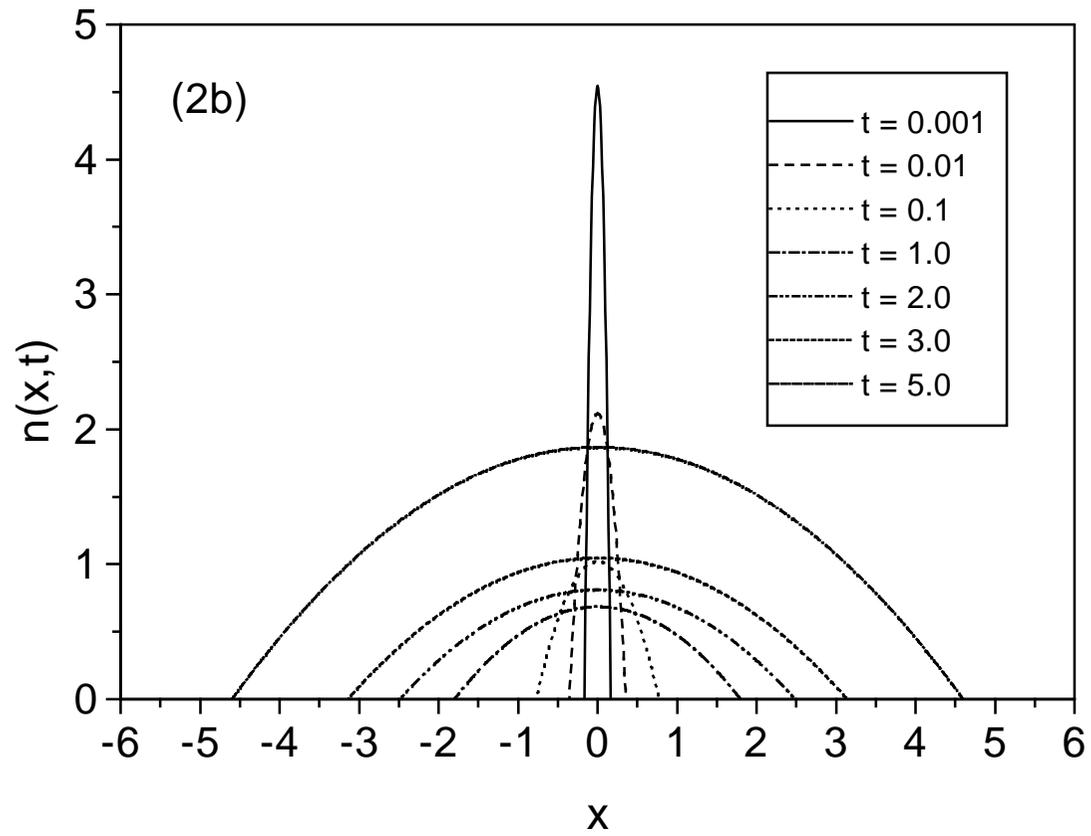
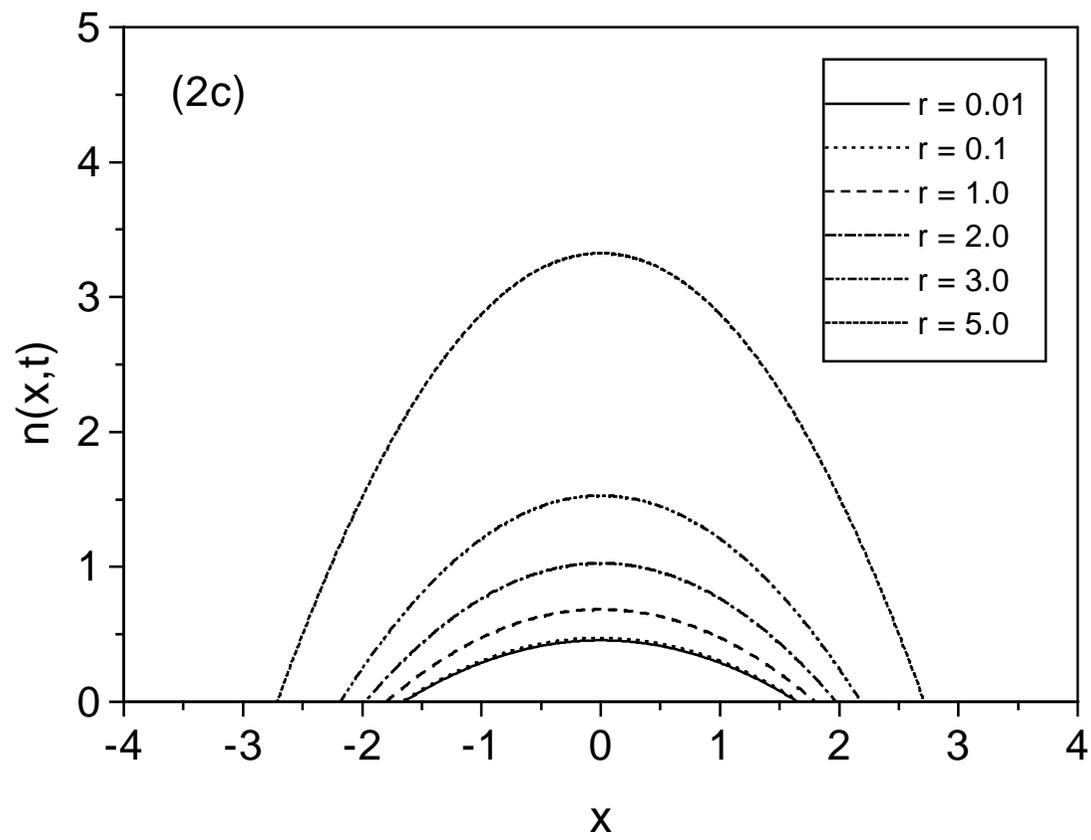
\end{document}